\documentclass[prd, onecolumn,nofootinbib,notitlepage,floatfix]{revtex4-1}

\usepackage{amsmath}
\usepackage{graphicx}
\usepackage{dcolumn}
\usepackage{bm}
\usepackage{epsfig}
\usepackage{amssymb,latexsym,mathrsfs}
\usepackage{graphicx}
\usepackage{color}
\usepackage{hyperref}

\hypersetup{
    colorlinks=true,
    linkcolor=red,
    citecolor=blue,
} 

\newcommand{\be}{\begin{equation}}
\newcommand{\ee}{\end{equation}}
\newcommand{\bs}{\begin{split}} 
\newcommand{\bea}{\begin{eqnarray}}
\newcommand{\eea}{\end{eqnarray}}

\begin{document}

\title{Isogrowth Cosmology (and How to Map the Universe)} 

\author{Eric V.~Linder$^{1,2}$}
\affiliation{
$^1$Berkeley Center for Cosmological Physics \& Berkeley Lab, 
University of California, Berkeley, CA 94720, USA\\  
$^2$Energetic Cosmos Laboratory, Nazarbayev University, 
Nur-Sultan, 010000, Kazakhstan}

\begin{abstract} 
While general relativity ties together the cosmic expansion history 
and growth history of large scale structure, beyond the standard model 
these can have independent behaviors. We derive expressions for 
cosmologies with identical growth histories but different expansion
histories, or other deviations. This provides a relation for isogrowth 
cosmologies, but also highlights in general the need for observations to 
measure each of the growth, expansion, gravity, and dark matter 
property histories. 
\end{abstract}

\date{\today} 

\maketitle


\section{Introduction}

Mapping the expansion history of the universe, e.g.\ through 
supernova distances \cite{perlmutter,riess}, led to the discovery 
of cosmic acceleration, replacing the then standard model of 
a matter dominated universe. Within general relativity, the 
growth of linear density perturbations into large scale structure 
-- the cosmic growth history -- is determined by the expansion 
history (so long as the matter density behaves in the standard 
way). 

However, shortly after cosmic acceleration became part of 
the new standard model, the recognition that it could arise 
not only from an additional energy density component with 
strongly negative pressure but also from a modification of the 
gravitational sector severed the lockstep between expansion 
and growth histories \cite{lue,song,knox,lin05}. Cosmologies 
with identical expansion histories could have different growth 
histories, due to the differing gravity. In this sense the standard 
model has moved from having two types of cosmic history, with 
one determining the other, to possessing three cosmic histories, 
with any two determining the third. 

Here we turn the situation around and study the inverse 
case of cosmologies with identical growth histories, and derive 
the necessary relations for the expansion and gravity histories. 
Cosmic growth history can be mapped through observations 
such as redshift space distortions in galaxy clustering, and as 
a blend with expansion history (and gravity) in probes such as 
weak gravitational lensing. As the upcoming generation of 
experiments prepares to deliver such data, and planning is 
underway for the science cases of the next generation of 
experiments, it behooves us to study the relation -- and 
freedom -- between the different types of cosmic history. 
That is, what should be the vision of how to map the universe 
so as to truly understand our cosmology (cf.~\cite{hut15,intertwine})? 

Another element is the division between the matter whose 
clustering the observations measure, and other energy density 
components that do not cluster effectively. In one sense, this 
occurs at the background expansion level, where distances 
measure the Hubble expansion rate $H(a)$, but the division 
into matter and dark energy  is done separately. This is 
sometimes referred to as the ``dark degeneracy'' 
\cite{0702615,1509.08458}. Here we are more interested in 
the perturbative level: what enters as the matter density 
perturbation $\delta\rho$ in the 
Poisson equation, and hence the growth evolution equation -- 
i.e.\ what clusters. Having only a fraction of the nonrelativistic 
matter cluster would imply physics beyond the standard 
cosmological model, whether arising from interactions with another 
component or internal properties. Such ``dark matter property 
history'' can have interesting implications, and be constrained 
in turn by cosmic probes (see, e.g., \cite{lindm}). 

In Section~\ref{sec:gro} we derive the relations for 
isogrowth cosmology, and in Section~\ref{sec:cases} 
investigate three subcases where certain aspects of the 
physics have freedom while fixing others (i.e.\ the matter, 
gravity, and expansion behaviors respectively). We 
discuss the use of isogrowth cosmology as a clear 
demonstration of freedoms and connections, and conclude 
with the general vision of mapping all four histories, in 
Section~\ref{sec:concl}.

\section{Cosmic Growth History} \label{sec:gro} 

The growth of matter perturbations $\delta\equiv\delta\rho_m/\rho_m$
into large scale structure is given by
\be
g''+\left[4+\frac{1}{2}\left(\ln H^2\right)'\right]\,g'+\left[3+\frac{1}{2}\left(\ln H^2\right)'-\frac{3}{2}G_N G(a)\Omega_m^{\rm cl}(a)\right]\,g=0\,,
\label{eq:gro} 
\ee 
in the linear density regime on subhorizon scales, with prime
denoting $d/d\ln a$ for $a$ the cosmic scale factor (see, e.g., \cite{lin05}). 
Here $g(a)=[\delta(a)/a]/[\delta(a_i)/a_i]$ is
the normalized growth factor (equal to one during standard matter
domination), $H(a)$ is the Hubble parameter describing the expansion
history, $G_N$ is Newton's constant, $G(a)$ a dimensionless modification
of the gravitational coupling to matter perturbations, and
$\Omega_m^{\rm cl}(a)$ is the fraction of the critical density contributed
by clustered matter. In the standard model there is no distinction
between $\Omega_m(a)$ and $\Omega_m^{\rm cl}(a)$ since there all matter
(i.e.\ nonrelativistic, zero pressure energy density) clusters.
However here we want to allow the possibility that only part of the
matter clusters; we will abbreviate $\Omega(a)\equiv\Omega_m^{\rm cl}(a)$
since we will be adding subscripts labeling different cosmologies.

Thus we see that outside the standard model there is freedom to
loosen the connection between expansion $H(a)$ and growth $g(a)$.
For example, at the background expansion level one can define an
effective energy density corresponding to gravitational modifications that
will exactly match the contribution of a physical energy density to the
Hubble parameter, giving identical $H(a)$; however the gravitational
modifications will enter in $G(a)$ and change the growth. Here we
consider identical growth histories $g(a)$, i.e.\ $\delta(a)$, and
derive the different expansion histories.

Two pedagogical points: The $G(a)$ here corresponds to $G_{\rm matter}(a)$, also
sometimes called $\mu(a)$, the gravitational coupling strength to
matter that appears in the modified Poisson equation for matter.
There is further gravitational freedom beyond this, which does not
enter here. Second, one could ask for identical growth history in
terms of $\delta(t)$ rather than $\delta(a)$. This is less interesting
in that we measure matter clustering at different redshifts, not
times, and if we allow differing expansion histories then the $a(t)$
relation is not fixed. 

Considering two different cosmologies, 1 and 2, with identical growth
histories $g(a)$, and hence $g'$ and $g''$, Eq.~\eqref{eq:gro} yields
the matching condition
\be
G_2(a)=G_1(a)\,\frac{\Omega_1(a)}{\Omega_2(a)}-\frac{f(a)}{3\Omega_2(a)}\,
\left(\ln \frac{H_1^2}{H_2^2}\right)'\,, \label{eq:match} 
\ee
where we have used that the growth rate $f(a)\equiv\delta'/\delta=1+g'/g$. Note that
the growth rate enters observationally in redshift space distortions.
Equation~\eqref{eq:match} shows the condition needed on the gravitational
coupling modification for the growth histories to match between the
two cosmologies. One could adopt cosmology 1 as being a 
general relativity standard cosmology, say, in which case $G_1(a)=1$,
$\Omega_1(a)=\Omega_m(a)=\Omega_{m,0}a^{-3}/[H_1(a)/H_1(a=1)]^2$.

We can write the matching condition needed on the expansion history
instead, obtaining 
\be
H_2(a)=H_1(a)\,\frac{H_2(a=1)}{H_1(a=1)}\ 
e^{(3/2)\int_0^{\ln a}d\ln x\,[G_2(x)\Omega_2(x)-G_1(x)\Omega_1(x)]/f(x)}\ .
\label{eq:matchh}
\ee 

Either Eq.~\eqref{eq:match} or Eq.~\eqref{eq:matchh} demonstrate that 
isogrowth cosmologies impose a relation among expansion history, 
gravity history, and matter clustering, while allowing freedom to trade 
between them. This is unlike general relativity standard cosmology where 
isogrowth determines exactly the expansion history (and vice versa).

\section{Three Cases} \label{sec:cases} 

Given the general relation, we can describe three special cases where 
we have both identical growth histories and one other type of history.

\subsection{Identical Growth and Clustering Matter} \label{sec:idcl} 

In standard cosmology all the nonrelativistic matter clusters, 
$\Omega(a)=\Omega_m(a)$. If we preserve this, then we still retain 
the freedom to trade off a modified expansion vs a modified gravity. 
That is, identical growth will result when either of the two following 
equivalent expressions holds 
\bea  
G_2(a)&=&G_1(a)-\frac{f(a)}{3\Omega_m(a)}\,
\left(\ln \frac{H_1^2}{H_2^2}\right)' \label{eq:matchcl}\\ 
H_2(a)&=&H_1(a)\,\frac{H_2(a=1)}{H_1(a=1)}\ 
e^{(3/2)\int_0^{\ln a}d\ln x\,[G_2(x)-G_1(x)]\Omega_m(x)/f(x)}\ .
\label{eq:matchhcl} 
\eea 

It is interesting to write the expansion derivative in terms of 
the total background equation of state $w(a)$ for the respective 
cosmology, giving 
\be 
G_2(a)=G_1(a)-\frac{f(a)}{\Omega_m(a)}\,[w_2(a)-w_1(a)]\ . 
\ee 
This ties together key cosmological quantities of gravity, 
matter density, and equation of state histories.

\subsection{Identical Growth and Gravity} \label{sec:idgrav} 

If we preserve general relativity, so that $G(a)=1$, then we 
can still obtain identical growth with different expansion histories 
if we adapt the matter clustering. This could be done through a 
certain fraction of the dark matter, say, interacting with another 
energy density component, or a self interaction. If the full matter density does not 
cluster, this can affect the initial conditions for the growth equation, 
introduce an extra integrated Sachs-Wolfe effect in the cosmic microwave 
background (CMB), and generically affect redshift space distortion 
observations (except here we are fixing the growth rate). See 
\cite{lindm} for discussion concerning the interplay of effects, 
and some experimental constraints. 

The matching conditions are 
\bea 
\Omega_2(a)&=&\Omega_1(a)-\frac{f(a)}{3}\,
\left(\ln \frac{H_1^2}{H_2^2}\right)' \label{eq:matchgr}\\ 
H_2(a)&=&H_1(a)\,\frac{H_2(a=1)}{H_1(a=1)}\ 
e^{(3/2)\int_0^{\ln a}d\ln x\,[\Omega_2(x)-\Omega_1(x)]/f(x)}\ ,
\label{eq:matchhgr} 
\eea 
equivalently. 
Note that if we take matter to cluster in the standard way, 
so $\Omega_2(a)=\Omega_1(a)$, then 
since we are in general relativity this means that isogrowth cosmologies 
would imply isoexpansion cosmologies. (In fact, this holds as 
well with $G_2(a)=G_1(a)\ne 1$, i.e.\ the same gravitational 
cosmologies even if the gravity is not general relativity, so long 
as the form of the density perturbation growth equation is 
not altered.)

\subsection{Identical Growth and Expansion} \label{sec:idexp} 

The last case is a curious one, where the two main cosmic histories 
remain identical, in particular the expansion history $H(a)$ 
in addition to growth, yet there is freedom distinct from the 
standard model. The matching condition is 
\be  
G_2(a)\,\Omega_2(a)=G_1(a)\,\Omega_1(a)\,. \label{eq:matche} 
\ee 
That is, the effective gravitating clustering matter densities $G\Omega$ must 
be identical, but we can trade off the gravitational coupling 
strength against the clustering matter density. For example, 
enhanced gravitational strength can give the same growth even 
with a smaller clustering matter density. 

Of course this freedom is not available within general relativity, 
where $G_2=G_1=1$. Nor may we do a simple change in 
the present matter density assuming all the matter clusters,  
$\Omega_{m2,0}\ne\Omega_{m1,0}$ since in the matter dominated 
epoch the expansion goes as 
$H^2=H_0^2\Omega_{mi,0}a^{-3}$ for $i=1$, 2 so for the same 
expansion one must have $\Omega_{m2,0}=\Omega_{m1,0}$.

\section{Conclusions} \label{sec:concl} 

Just as observations determining the expansion history alone are 
not sufficient to characterize the cosmology, neither is growth history 
data alone. We presented explicit expressions for isogrowth 
cosmologies, where variations in some set of expansion, gravity, 
and matter clustering histories could still yield identical growth 
history. 

However, isogrowth imposes relations among the three other 
histories and if we further adopt one of those to remain identical 
when varying cosmology, we obtain three subcases. The first, 
with all of the matter clustering, keeps the matter sector as standard, 
and trades off the gravity and expansion histories. The second, 
with gravity unchanged (e.g.\ standard general relativity) shows 
that variations in the dark matter sector can compensate for 
a differing expansion history. The last, 
where both growth and expansion are preserved, highlights the 
important role of the effective gravitating clustering matter 
density. 

Thus, to truly understand our cosmology we must measure 
at least three of the growth, expansion, gravity, and dark matter 
property histories, while mapping the fourth serves as an important 
crosscheck. Note that there are further aspects of gravity 
not entering in the discussion here: gravity affects not only 
matter but light, and the quantity $G_{\rm light}(a)$ (sometimes called 
$\Sigma$) entering the modified Poisson equation for light 
affects gravitational lensing and other light propagation. 
Gravitational waves also probe gravity in a somewhat 
distinct manner, e.g.\ through the running of the effective 
Planck mass. Furthermore there are new aspects that 
enter beyond the linear density regime, including scale 
dependence. 

The mapping of all these histories through cosmological 
observations should be a central part of the vision for 
understanding our universe in the next two decades.

\acknowledgments  

This work is supported in part by the Energetic Cosmos Laboratory, and
by the U.S.\ Department of Energy, Office of Science, Office of High Energy 
Physics, under contract no.\ DE-AC02-05CH11231.

\end{document}